# Fake or Credible? Towards Designing Services to Support Users' Credibility Assessment of News Content

Enrico Bunde
Ruhr-Universität Bochum
enrico.bunde@rub.de

Niklas Kühl
Karlsruhe Institute of Technology
kuehl@kit.edu

Christian Meske
Ruhr-Universität Bochum
christian.meske@rub.de

**Abstract**

*Fake news has become omnipresent in digitalized areas such as social media platforms. While being disseminated online, it also poses a threat to individuals and societies offline, for example, in the context of democratic elections. Research and practice have investigated the detection of fake news with behavioral science or method-related perspectives. However, to date, we lack design knowledge on presenting fake news warnings to users to support their individual news credibility assessment. We present the journey through the first design cycle on developing a fake news detection service focusing on the user interface design. The design is grounded in concepts from the field of source credibility theory and instantiated in a prototype that was qualitatively evaluated. The 13 participants communicated their interest in a lightweight application that aids in the news credibility assessment and rated the design features as useful as well as desirable.*

## 1. Introduction

Deception in the form of fake news is an omnipresent phenomenon in our digitalized world [1]. The swift dissemination of false information is reinforced through social media platforms' wide adoption and usage [2]. Today, these social media platforms have developed into a widespread source to consume as well as share news [3]. A significant difference compared to traditional media persists in the underlying algorithms that provide targeted information for the user without being transparent [4]. Some actors on social media platforms produce intentionally misleading articles and try to resemble legitimate news organizations, while satirical sites publish news, which may be perceived as facts by the readers [5].

We understand fake news as "[…] news articles that are intentionally and verifiably false and could mislead readers […]" ([5], p. 213). Deception in general and fake news specifically bear the far-reaching potential risk that humans take actions based on them. For example, fake news can influence the outcomes of democratic elections [5, 6]. Another example refers to the COVID-19 pandemic. In 2020 it was claimed that there is a link between 5G and the health crisis. Consequently, 5G radio masts in England and Northern Ireland were set on fire [7]. Therefore, it is crucial to enable internet users to identify potential fake news and support them in assessing the credibility of news content [1, 3]. Information systems research plays a significant role in providing solutions to that problem as the reorganization and adjustment of information systems (e.g., social media platforms) plays a significant role in reducing the impact of false information [8].

The subject of fake news is examined from various perspectives using different approaches. For example, behavioral research has investigated the effects of user and expert reputation ratings in the context of fake news interventions [2, 4, 8]. Another subject that is gaining in relevance is the automated classification of fake news using data mining approaches from the field of artificial intelligence, such as machine learning or deep learning [9-11]. In future research, we will develop a service artifact based on artificial intelligence since it is described as effective [12]. In this manuscript, we focus on the design knowledge for user interfaces for services that support the credibility assessment of news content. Consequently, we establish the following research question: *Which design features should user interfaces for services integrate to support users in assessing the credibility of online news content and their source?*

To answer this research question, we conduct a design science research (DSR) project [13, 14]. The aim is to introduce novel design knowledge [15]. In doing so, we derive design principles as well as design features that are instantiated in a first prototype and qualitatively evaluated with 13 participants.

The remainder of this manuscript is structured as follows. In the next section, we describe the DSR approach, which represents the methodical basis. We then start with conducting the relevance cycle, highlighting the problem statement. This is followed by the rigor cycle, where we describe different nuances of fake news and corresponding risks. Additionally, major challenges are derived from literature which are then addressed by design requirements. Subsequently, we conduct one design cycle to establish, implement and evaluate first design knowledge. This is followed by a discussion, after which we conclude this manuscript.

## 2. Design Science Research-Based Methodology

As an overall research design, we choose DSR and follow the methodology proposed by Hevner and Chatterjee [14]. They suggest that a DSR project should cover the cycles of investigation, the relevance cycle (targeting the practical problem within the naturalistic application domain), a rigor cycle (focusing on the knowledge base and foundations), as well as one or multiple design cycles (building and evaluating the research artifact). Figure 1 illustrates this DSR approach for the work at hand.

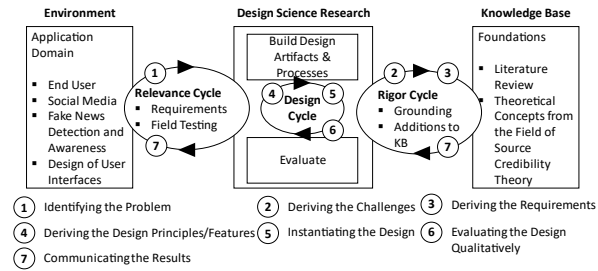

**Figure 1.** DSR approach (based on [14]).

For the evaluation, we follow the framework for evaluation in design science research (FEDS) proposed by Venable et al. [16] with one design cycle (DC1) as depicted in Figure 2. Our design is further motivated and grounded in theoretical concepts related to the source credibility theory (SCT) [17, 18]. We argue that SCT is a suitable theoretical foundation for the design. Source credibility itself represents an essential aspect of automated fake news detection [10]. Moreover, SCT was utilized to investigate the effect of source credibility on information system acceptance and its influence on cognitive responses of users [17, 19, 27]. In addition, the SCT will serve as the foundation to develop a research model for a quantitative evaluation in future research (e.g., an online experiment in design cycle two (DC2)). Furthermore, in DC2 we will implement a machine learning-based model for fake news detection and integrate it with the user interface (e.g., by providing a browser plugin or web-based application).

In the initial design cycle, we further analyze the problem statement of the relevance cycle and derive insights to formalize the knowledge into challenges that are mapped to precise design requirements (DRs). We then propose a set of initial design principles (DPs) that are further translated into concrete design features (DFs) that are instantiated in a prototypical user interface for a service that supports the credibility assessment. The initial design was evaluated qualitatively through open-ended interviews with 13 participants (5 females, 8 males; age $M$ = 33.5 years, $SD$ = 9.8; interview duration: $M$ = 19.2 minutes, $SD$ = 2.8; daily social media usage: $M$ = 1.9 hours, $SD$ = 1.3). They were recruited via snowball sampling [20]. All participants use multiple social media platforms (e.g., Twitter, Instagram, or Facebook), and all participants stated they had encountered fake news or untruthful news content on social media platforms. Within the qualitative evaluation, we focused on the dimensions of validity, utility, and efficacy [16, 21].

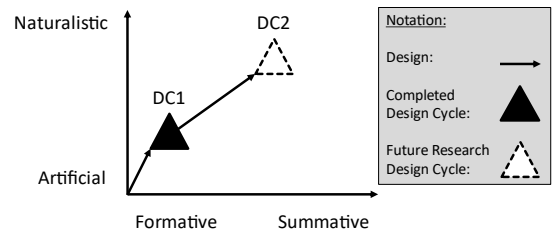

**Figure 2.** DSR evaluation strategy (based on [16]).

## 3. Relevance Cycle: The Potential Danger of Fake News

Within the relevance cycle, we identify the challenges of the application domain that need to be addressed by the DSR project [21, 22]. A precise problem formulation, characterization, and positioning into a problem space is a crucial factor for DSR research [14, 15, 22]. Therefore, this section highlights the dangers of deceptive information, especially fake news, their consequences, and the lacking design knowledge within this research area.

There are many different actions and threats in the real world, which can be traced back to fake news and have been reported in the literature. A significant focus of research in the context of fake news is politics, often in connection with social media platforms such as Facebook or Twitter (e.g., [5, 8, 23]). In the context of the 2016 presidential election in the U.S., it was found that 1% of Twitter individuals accounted for the exposure of 80% of fake news, and additionally, only 0.1% were accounted for sharing nearly 80% of fake news [6]. Research with a focus on Twitter has also shown that false information in many cases is spread (i.e., retweeted) much faster and further by users than accurate information [1, 24]. The danger can be even more severe if the dissemination of false information is used by terrorist organizations, which was already documented when individuals from Europe joined ISIS and their training for terrorist acts [25].

In literature, several vital contributions in the context of fake news can be found. On the one hand, with a behavioral or social science focus such as the presence of fake news flags and their influence on a user's cognition as well as judgment or the impact on political elections (e.g., [1, 2, 4-6]). On the other hand, with a more technical focus regarding the automated detection or the application of neural networks (e.g., [3, 9, 11, 12, 26]). These research contributions influence the established design knowledge in this manuscript (see rigor cycle). However, despite these valuable contributions, there is a gap of clear and applicable design knowledge in the context of fake news detection services. The following section presents related work, identifies challenges, and contributions in existing literature, which build the foundation for the DRs, DPs, and DFs.

## 4. Rigor Cycle: Related Work

The aim of the rigor cycle is to access and provide knowledge to the DSR project, as well as to guarantee that the created design depicts a clear research contribution [14, 22]. Ultimately, prior knowledge that will be used in the DSR project is presented, also denoted as input knowledge [15].

### 4.1 Nuances of Fake News, Actors, and Motivations

There are different nuances of fake news on the web and on social media platforms [3, 9]. One complex of themes, which is a recurring phenomenon, are conspiracy theories [8]. These types of theories exist in a variety of subject areas [5]. Additional types of content that can contain false information are, for instance, fabricated information, hoaxes, propaganda, photo manipulation, or clickbait [25, 27]. Another related concept is truth discovery which aims to detect true facts from various conflicting sources [28]. Fake news detection can benefit from different aspects of truth discovery techniques as it aims to identify the credibility of the source as well as the object's truthfulness alike [9, 28]. The phenomenon of fake news has led to an increased interest in fake news, which can be traced back to a widespread impact on public opinions and events [12]. Within social media platforms, different actors can have an interest in the propagation of fake, deceptive, or false information. For example, activists, political organizations, governments, criminals, journalists, or so-called trolls may have an interest to deceive or misinform individuals [25]. Just as different actors can stand behind fake news, the motives can also be diverse, including malicious intentions, the will to gain influence, profit, or fun [10, 25].

Different risks emerge within this context. A specific risk with respect to social media platforms is the speed at which information, including fake news, is shared, and therefore spread [4]. Another danger of news on social media platforms lies in the blurring of information sources and users having to navigate through various details to identify an individual source [27]. An additional problem that emerges in social media platforms is the echo chamber effect [29]. Social media platforms present information and opinions that users potentially agree with, consequently leading to a filter bubble over time [8, 9].

### 4.2 Detection of Fake News with Artificial Intelligence

The rapid dissemination of fake news and its potentially negative impact on democracy, justice, or society is increasing the demand for detection as well as intervention services for fake news [10]. The automated detection of such news remains a challenge due to the unique characteristics it possesses [9, 26]. Nevertheless, there is active research on this subject in the field of artificial intelligence using machine learning or deep learning, with effective models being developed [12]. Scholars investigate fake news detection from different perspectives, integrating different features, and utilizing a variety of methods such as machine learning models (e.g., support vector machines or random forests) or deep learning models (e.g., convolutional neural networks or recurrent neural networks) [3, 12, 26].

These approaches can process a wide variety of attribute types and features, for example, attributes related to quantity with features such as the number of characters, words, sentences, or paragraphs [10]. In addition, machine learning-based fake news detection can integrate further data related to the news content as well as the social context [9]. For example, the topic distribution or sentiment information are part of content-related features whereas the follower-friend ratio or number of friends are part of social context-related features [3, 12, 17]. Moreover, machine learning and deep learning approaches are able to process mixed data input types such as texts in combination with images [3]. Such sophisticated approaches are becoming increasingly performant in that they can achieve higher performance, more quickly detect fake news, or require less data for effective training [11]. Despite these positive developments, there are still many exciting research opportunities, which relate, for example, to the early detection of fake news (i.e., the detection before the content spreads) or explainable fake news detection (i.e., the explainability of the underlying machine learning or deep learning approaches) [10].

## 4.3 Challenges and Contributions within Research on Fake News Detection

As described in the previous sections, various contributions to the knowledge base regarding fake news can be found across the literature. Nevertheless, explicit design knowledge for user interfaces of services that offer support in the credibility assessment of online news content, to the best of our knowledge, is nonexistent. Table 1 summarizes the six main challenges (*C1-C6*) that were synthesized. They serve as the basis for the derived DRs and, in consequence, the established design knowledge.

One problem in the presentation of news in social media platforms lies in the underlying algorithms that are used to compile the news, content, and that introduce a transparency problem regarding the source quality as well as credibility [1, 3, 8]. Moreover, research emphasized that the credibility of the source and news should be distinguished [17, 27] which is made difficult through the transparency problem (*C1*). Within social media platforms, the news is automatically composed through algorithms, and the user's feed is filled with short snippets or previews that represent the linked content [2]. The core problem lies in the fact that the source can often not be recognized easily and is therefore overlooked since it is rarely highlighted [4, 10]. This is problematic as it complicates the source credibility assessment [17] (*C2*). The dissemination of fake news is facilitated by the structure of social media platforms [12, 26]. Research has uncovered that fake news can spread significantly faster and more broadly on social media platforms than true news [11, 24]. These factors make it increasingly important to detect fake news at the earliest possible stage, ideally, before it has been widely disseminated [3, 11] (*C3*). The automated combination and recommendation of content such as news articles is often based on automated processes, for example, with machine learning or deep learning [9, 11, 26]. Due to their potentially high degree of adaptation, there is a risk that users get only content suggested, which is more likely to be believable, resulting in a so-called filter bubble [5, 9, 23] (*C4*). Another challenge lies in the confirmation bias, which "[…] connotes the seeking or interpreting of evidence in ways that are partial to existing beliefs, expectations, or a hypothesis in hand." ([30], p. 175). In this context, people can be selective regarding the evidence that they seek. Therefore, the confirmation bias is a risk, which can be further strengthened when combined with a filter bubble [4, 9, 29] (*C5*). Moreover, people tend to believe and trust information that supports their existing ideas and beliefs, which leads to the fact that content and news representing their own ideas or beliefs are less scrutinized [1, 2, 4, 29, 30] (*C6*).

**Table 1.** Identified challenges and contributions in the context of fake news detection.

| C | Description | Source |
| --- | --- | --- |
| C1 | Complex algorithmic models are used to predict and maximize engagement with the provided content. Hence, the influence of the users on the presented content and their sources decreases. | [1, 3, 8, 17, 27] |
| C2 | Through the processing and intermixing of different sources as well as the sharing, spreading, and presentation of content as snippet, sources are hard to identify. Users must navigate through various details to identify a source. | [2, 26, 17, 27] |
| C3 | The structures of social media platforms favor the easy publication and rapid as well as wide dissemination of fake news, which is often shared more widely than true news. | [3, 11, 24] |
| C4 | Due to the automated recommendation of the content and a potentially high degree of adaptation, there is a risk of a filter bubble, meaning that only content that a user is more likely to believe is suggested. Furthermore, the user is not provided with adequate information or sources to evaluate the credibility or truthfulness of the provided content. | [5, 9, 23, 29] |
| C5 | The combination of a confirmation bias and filter bubble is a potential risk. Since users are more likely to believe the news that comply with existing beliefs and within a filter bubble, more of such content could be suggested. | [4, 9] |
| C6 | People often tend to believe and trust information that supports existing ideas or beliefs, and hence, the content is less likely to be scrutinized. | [1, 2, 30] |

# 5. Designing User Interfaces for Fake News Detection Services

To design our artifact, we derive the DRs, followed by DPs and DFs. Afterward, we demonstrate our initial prototypical user interface for services that support the credibility assessment, the qualitative evaluation as well as the results. We ground the design knowledge within theoretical concepts of the SCT.

## 5.1 Deriving Design Requirements

First, we derive the DRs and map them to the corresponding challenges that were identified in the rigor cycle above. Table 2 summarizes the DRs.

Algorithmic recommendations lead to the content which the user is presented with [1, 8]. Since the user has no influence on this, the credibility of the sources should be communicated. In this context, source credibility can influence the extent to which users are accessing the source to gain expertise or knowledge [17]. Moreover, source credibility can affect judgments [31] and can influence an users' perception of source as well as the news content [18]. Consequently, the artifact should integrate signals to present the source credibility so that users are supported in credibility assessment. Hence, we establish *DR1*.

Through the presentation of news on social media platforms as a snippet or preview, as well as the processing and intermixing of different sources, these sources can be overlooked [2, 26, 27]. If users cannot identify the source, they cannot assess the credibility and may feel uncertain about the content presented to them [17-19]. Moreover, in the context of review websites, scholars have described that decision outcomes can be influenced by the perceived credibility of a source [31]. Consequently, the artifact should present the source of the content so that users can assess the source credibility. Hence, we establish *DR2*.

There are different approaches to detect fake news, such as feature-oriented (e.g., with a focus on news content and the social context) or model-oriented (e.g., utilizing semi-supervised, supervised, and unsupervised learning approaches) [9]. More recently, artificial intelligence and, more precisely, machine learning as well as deep learning have been utilized for detecting fake news with high effectiveness [12]. Different approaches already achieve high performance [3, 10, 11]. Results of these classifications must be communicated in an easy-to-understand way. This should be realized to support the users' credibility assessment [12, 17, 23]. Consequently, the artifact should clearly communicate the machine learning-based classification outcome supplemented by the news content. Hence, we establish *DR3*.

Due to the automated recommendation of content, there is a high degree of adaptation (e.g., personalization) that leads to the risk of a filter bubble [5, 9, 23]. Additionally, people tend to have a confirmation bias which is a crucial factor in the context of fake news detection and awareness [4, 8, 9]. Moreover, this is problematic as users are less confronted with opinion-challenging information and content [29]. Consequently, the artifact should offer other topics to break through a potential filter bubble. Hence, we establish *DR4*.

Furthermore, people often tend to believe and trust information that supports their existing beliefs [1, 2, 4, 30]. Consequently, the artifact should offer information regarding counter indicative facts so that users can deal with various sources and arguments on the present news content. Hence, we establish *DR5*.

**Table 2.** Mapping challenges to design requirements.

| Mapping | | Description |
|---|---|---|
| C1 | DR1 | The artifact should integrate signals that describe the source credibility of the presented content. |
| C2 | DR2 | The artifact should present the source of the content so that it cannot be overlooked. |
| C3 | DR3 | The artifact should clearly communicate the machine learning-based classification outcome supplemented by the news content. |
| C4; C5 | DR4 | The artifact should offer other topics to break through a potential filter bubble. |
| C6 | DR5 | The artifact should offer information regarding counter indicative facts when false, fake, or deceptive content is identified. |

## 5.2 Deriving Design Principles and Design Features

The derived DRs are the basis for the initial set of DPs and corresponding DFs. Table 3 summarizes the DPs and illustrates the relationship between the DRs and DPs. In the following, we derive the DPs.

There is a link between the source credibility and the credibility of the provided news [18]. Source credibility consists of different dimensions such as trustworthiness or expertise [17]. The credibility of the source is another important element and feature, which could be based on user or expert ratings [2, 8]. Research has uncovered high source ratings lead to users do not take much effort on critical thinking, though lower ratings lead to users paying more attention to the rating and its mechanism [8]. Moreover, the emphasis on the credibility of the presented information should be enhanced [1]. Since we aim to empower the user to assess the credibility of news content, the design should integrate the original source and the source credibility. Consequently, the artifact should present the source and visualize the source credibility. Hence, we establish *DP1*.

The assessment of the credibility of news articles is frequently tackled with approaches from the field of data mining or artificial intelligence [3, 9-12]. These approaches are well established in different contexts, such as social media analytics, where for instance, deep learning can be utilized to classify the content using different features such as visual or textual content [32]. The classification of content such as fake news is challenging. Nevertheless, deep learning can process a multitude of different features to reach high performance in this task [26]. In the context of recommender systems, research indicates that users prefer the communication of a prediction on different forms of rating scales [33]. Lastly, the design should provide supplemental information in combination with the content of a user's feed [2]. Hence the original news content must be provided, and the design should be unobtrusive. Consequently, the artifact should clearly communicate the classification outcome in a visual rating scale and provide the piece of content. Hence, we establish *DP2*.

The concepts of an echo chamber and filter bubble are also described as information-limiting environments, where social networks constrain information sources that could shield users from opinion-challenging information [29]. This can also lead to the circumstance that users form groups with like-minded users and, for example, polarize their opinions [3, 9]. Another consequence could be the lack of ideological diversity of news and information sources when these contents are discovered through friendships on social media platforms [29]. Therefore, we want to provide the user with an opportunity to break out of their potential filter bubble by providing other topics to explore. Such a feature could also be implemented with data visualization techniques and is able to increase the user experience and trust in systems [23]. Additionally, when a present news article is classified as potentially fake, the user should be provided with sources that underpin this outcome, for example, sources of disproof, curated by other users, experts, or fact-checking services [3-5, 8, 11]. These sources of disproof enable the user to evaluate the counter indicative contents as well as its source, which can be important for the credibility assessment [17, 18]. Consequently, the artifact should provide an opportunity to discover other topics as well as sources of disproof in case of potential fake news detection. Hence, we establish *DP3*.

**Table 3.** Mapping design requirements to design principles.

| Mapping | | Description |
|---|---|---|
| DR1; DR2 | DP1 | The artifact should present the source and visualize the source credibility in a visual rating scale to support the source credibility assessment. |
| DR3 | DP2 | The artifact should clearly communicate the classification outcome in a visual rating scale and provide the piece of content to provide unobtrusive, designed decision support. |
| DR4; DR5 | DP3 | The artifact should provide an opportunity to discover other topics as well as sources of disproof in case of a potential fake news detection to break through filter bubbles and investigate the sources. |

The derived DPs are the basis for the initial set of concrete DFs that are instantiated within a first prototypical user interface for services that support the credibility assessment of news content [34]. Table 4 summarizes the DFs and illustrates to which DPs they can be mapped.

Research has shown that sources of news content in social media can easily be overlooked due to the presentation within the user feed and emphasized the important role of the source for the credibility assessment [18, 26, 27]. Consequently, the original source should be clearly positioned on top of the screen *DF1*. Moreover, the user interface should visualize the credibility of the source, which could be generated by fact-checking services, users, through data mining approaches or collaborative reputation systems [3, 9, 33, 35]. Consequently, the source credibility should be visualized as a rating scale *DF2*. The outcome of the prediction for the present news article will be presented as a rating scale [12, 33]. Consequently, the likelihood of the machine learning-based classification should be visualized as a rating scale *DF3*. The overarching goal of the design is to support the identification of potential fake news and create awareness. Therefore, the DFs should be supplemental and provided in combination with the content in a user's feed [2]. Consequently, the processed news article should be provided *DF4*. By providing other topics to the user, we aim to support them to break through a potential filter bubble and provide them with diverse information sources [23, 29]. Consequently, a feature to discover other topics should be provided *DF5*. When the design communicates a potential fake news article, sources of disproof should be provided so that users can evaluate the counter indicative contents [17, 18]. Consequently, sources of disproof should be provided if fake content is detected *DF6*.

Table 4. Mapping design principles to design features.

| Mapping | | Description |
|---|---|---|
| DP1 | DF1 | Provide the original source at the top of the screen. |
| | DF2 | Provide the source credibility as a visualized rating scale. |
| DP2 | DF3 | Provide the likelihood of the classification as a visualized rating scale. |
| | DF4 | Provide the processed news article. |
| DP3 | DF5 | Provide a feature to discover other topics. |
| | DF6 | Provide sources of disproof, if necessary. |

## 5.3 Prototypical Design of User Interfaces for Services that Support the Credibility Assessment

The following Figure 3 shows an example of the instantiated DFs for a credible news scenario. The implemented DFs are emphasized. The original source and the source credibility which is visualized as source credibility rating scale in the form of shield icons at the top address *DF1* as well as *DF2*. The more filled shield icons are presented, the higher the source credibility. At the bottom, the likelihood for the presence of fake news is given. Here, a visualized rating scale in the form of a traffic light is operationalized since we assume this DF can be understood very intuitively by users. For example, a red light indicates a high likelihood for fake news and a green light for a low likelihood. This addresses *DF3*. By providing the original news content as the main component, we address *DF4*. Another integral feature of the proposed design is to provide the users the opportunity to discover new topics and hence, break out of a potential filter bubble *DF5*. The design component should provide sources of disproof, which is especially relevant for content with a high fake news likelihood. These sources of disproof could be contributed by fact-checking services or even from other users and addresses *DF6*.

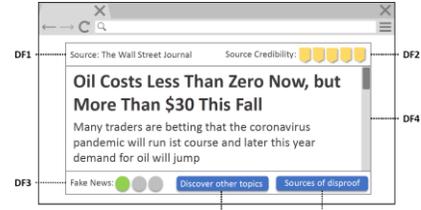

Figure 3. Prototypical user interface for a genuine (true) news article.

## 5.4 Qualitative Evaluation of the User Interface

The overarching objective of this evaluation was to confront as well as falsify the initial design for the user interface with real-world users. We have conducted 13 standardized open-ended interviews that followed the same semi-structured guidelines with the intention of taking each interviewee through the same evaluation procedure [20]. Prior to the interview, we introduced the design, prepared ten examples, and described the underlying mechanisms. We integrated questions with the intent to assess to which degree the DFs can achieve the overarching goals of the individual DPs. The interviews were conducted through Skype. The recording of the interview was followed by a transcription. The resulting data was evaluated following the thematic analysis as described by Braun and Clarke [36]. Table 5 provides an overview of the results that represent the underlying goal of each DP, as well as representative and recurring statements of the interviewees.

The *DP1 (DF1, DF2)* was described as a basic and essential feature from the user perspective. Participants indicated they want to be provided with the original source as one information input for their individual credibility assessment. In addition, some interviewees would participate in rating news sources on social media platforms to provide user-generated rating scales. With respect to *DP2 (DF3, DF4)*, participants liked the lightweight and unobtrusive DFs. They were rather deterred by the idea of using an additional service or external information system. Moreover, the fake news traffic light was described as an easy-to-understand classification outcome. The last *DP3 (DF5, DF6)* was also perceived as valuable. Some participants were extremely interested in the idea of discovering other topics outside of their individual user feed. They mentioned how recommendations within the user feed could evolve into boring or monotonous content. Moreover, they communicated interest in the sources of disproof. If the user is confronted with a classification that indicates that the present news content is fake or untruthful, they are willing to consciously look through these sources.

The evaluation also uncovered interesting insights for further development of the design. Many participants mentioned, if they are provided with automated classification, they are interested in reasons that have led to this outcome. Moreover, it was clearly communicated that the users would like to interact with the user interface in a real-world environment. Here, they would prefer a lightweight application with an unobtrusive design over an external information system or an additional service.

Table 5. Summary of recurring statements and feedback from the qualitative evaluation.

| DP | Goal | Representative and Recurring Statements |
|---|---|---|
| DP1 | Support the user's source credibility assessment process. | Alpha: "The original source is very important for me to individually evaluate the truthfulness of the outlet." Beta: "I like the provided rating for the source credibility and if available, I would rate sources if these ratings were curated by users." |
| DP2 | Support the user's assessment of a news' credibility with an unobtrusive design. | Gamma: "The design elements are intuitive, especially the fake news traffic light is an easy-to-understand indicator." Delta: "For me, the design elements are self-explanatory." Epsilon: "I like the idea of having a user interface extension rather than using a completely new system or service to assess the truthfulness of news content in my feed." |
| DP3 | Discover content outside the filter bubble and provide sources of disproof. | Zeta: "I would like to discover other topics because sometimes the recommended content can be boring or monotonous." Eta: "Sources of disproof could be an important feature to enable the end-user in evaluating the classification as well as for forming an own opinion." |

## 6. Discussion

### 6.1 Summary of Findings and Implications

We have presented the status quo on nuances and dangers of fake news, its detection through service artifacts, their design, as well as challenges and contributions, which built the basis for the design knowledge. We further grounded the design in theoretical concepts from the field of SCT. Moreover, we derived DPs and associated DFs. The design was instantiated in a prototypical user interface for services that support the credibility assessment of news content and qualitatively evaluated with 13 participants.

In summary, the results indicate that the initial DFs of the user interface are perceived as useful for the assessment of a news' credibility. The interview participants described that they prefer a lightweight user interface design (e.g., provided as a browser plugin) over complex information systems or external services. When such systems provide automated fake news classifications, participants demand some form of justification or explanation. This is an important finding as artificial intelligence is more frequently used for fake news classification [3, 9, 11, 12], and this can be addressed by integrating methods from the field of explainable artificial intelligence to generate explanations for black-box systems [37]. Moreover, the subject of explainable fake news detection has already been recognized in recent studies [10].

The qualitative evaluation uncovered users are willing to rate the information sources, which is in line with research on rating scales, for example, in the context of recommendation agents [33]. Furthermore, the participants rated the source as a piece of essential information for the credibility assessment, which was also found to be true in the context of fake news and its effects on behavioral intentions towards an advertised brand [18]. The visualized rating scale for source credibility was also found to be relevant for opinion formation. This supports the findings of Hsieh and Li [31], who found that source credibility can affect judgments. In addition, source-based fake news detection and the source credibility assessment on news authors or publishers are established approaches [10].

By initiating this DSR project and design knowledge, we follow the call for new systems of safeguards against fake news [1]. Based on Gregor and Hevner [21], we argue that the contribution of this DSR project can be categorized as an improvement. We developed a new solution (i.e., design knowledge) to a known problem (i.e., fake news detection) that was grounded in the status quo of research and qualitatively evaluated.

### 6.2 Limitations and Future Research

This DSR project was conducted according to established guidelines from the DSR community (e.g., [13, 14, 16, 22]). Nevertheless, the project presented here has limitations, as do other research projects. First, the here presented evaluation episode focuses on the design of user interfaces for services that support the credibility assessment of news content. Therefore, we have mostly disregarded the technical details of classifying fake news and the process of curating other topics or sources of disproof. Second, the here presented user interface was instantiated prototypically. Hence, it was not as interactive as a real-world software artifact, and therefore the engagement, as well as the interaction with the user interface, was limited.

We plan to address these limitations in future research by refining the design knowledge. In doing so, we implement a machine learning model for the automated fake news detection such as a convolutional neural network, which was already proven to be effective [12]. Despite the high performance of convolutional neural networks, they suffer under the black box problem which leads to the circumstance that users cannot understand the reasons for the outcome [37]. By utilizing methods from the field of explainable artificial intelligence, we address this problem [37], generate explanations for the underlying convolutional neural networks' decision-making. Therefore, our future focus lies on explainable fake news detection [10]. This information is brought together in the user interface and made available interactively to the user. Moreover, the design knowledge will be instantiated in a real-world artifact (e.g., browser plugin). This artifact will be developed with established programming languages and technologies such as Python, Django, Scikit-learn, and Keras. Consequently, we will evaluate this matured software artifact quantitatively in a real-world environment. Therefore, we use the SCT-grounded design to generate empirical insights of the influence on, for example, decision outcomes [31] or the perceived source credibility [18] in the context of fake news detection.

### 7. Conclusion

In this manuscript, we identified challenges in literature on fake news detection and derived specific design knowledge for user interfaces for services that support the credibility assessment of online news content. To achieve this, a DSR paradigm was applied [14]. Within the relevance cycle, the problem formulation was stated, while the rigor cycle covered related work. The identified challenges represented the foundation for the derived DRs, based on which the DPs and DFs were developed. A first design cycle was conducted, through which we prototypically implemented the design knowledge. The instantiated user interface was evaluated qualitatively with 13 participants and revealed the positive perception as well as the usefulness of the design. The results further highlight the versatile research opportunities within the field of artificial intelligence-based services to fight fake news in our digitalized and connected world.